\documentstyle[graphicx,aps]{revtex}

\begin{document}

\title{Scattering properties of a cut-circle billiard waveguide with
two conical leads} 
\author{Kathrin Fuchss, Suhan Ree, and L.E. Reichl}
\address{Center for Studies in Statistical Mechanics and Complex
Systems\\ The University of Texas at Austin\\ Austin, Texas 78712}

\date{\today}

\maketitle

\begin{abstract}

We examine a two-dimensional electron waveguide with a cut-circle
cavity and conical leads. By considering Wigner delay times and the
Landauer-B\"uttiker conductance for this system, we probe the effects
of the closed billiard energy spectrum on scattering properties in the
limit of weakly coupled leads. We investigate how lead placement and
cavity shape affect these conductance and time delay spectra of the
waveguide.

\end{abstract}

\bigskip

\section{Introduction}

The scattering properties of ballistic electron waveguides have been
studied by a number of authors, but primarily in the semi-classical
regime, where resonances strongly overlap \cite{baranger},
\cite{rouvines}. Several experiments \cite{exps} have shown that the
shape of the waveguide cavity can strongly affect the conductance,
depending on whether the shape induces integrable or chaotic
motion~\cite{gursoy}. In this paper, we investigate the scattering
properties of an electron waveguide whose scattering cavity strongly
resembles that of a quantum dot. We investigate this system in a
regime where the resonances are well separated in energy, and we look
at the effect of distortions in shape and changes of lead placement,
which change the symmetry of the scattering cavity, on the behavior of
the scattering resonances.

\begin{figure}[h]
\begin{center}
\includegraphics[scale=1.0]{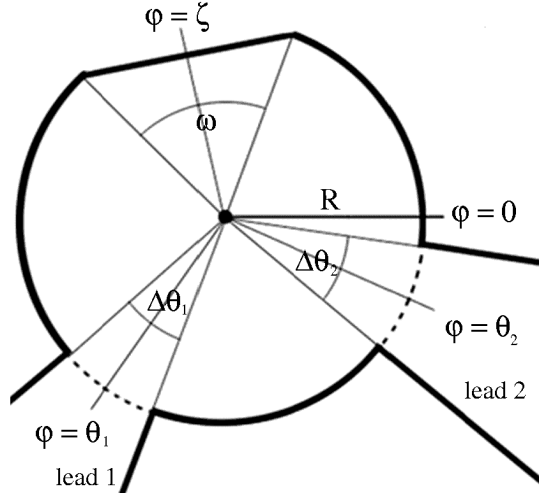}
\caption{Geometry of the scattering system with cavity radius $R$,
lead positions at $\theta_1$ and $\theta_2$, lead opening angles
$\Delta\theta_1$ and $\Delta\theta_2$, cut position at $\zeta$ and cut
angle $\omega$. \label{fig:geo}}
\end{center}
\end{figure}

The geometry of the waveguide we consider in this paper is shown in
Fig.~\ref{fig:geo}. The scattering system consists of a cavity with
two cone shaped leads attached, which we will treat as infinitely
long. The walls of both the cavity and the leads are assumed to be
infinitely hard (infinite potential) walls. The cavity has the shape
of a circle of radius $R$, but with a segment of the wall, subtended
by an angle $\omega$, replaced by a flat segment which we call the
cut. The opening angles of the leads are $\Delta \theta_1$ and $\Delta
\theta_2$. The angular positions of the center of the cut and the
center of the lead openings are denoted by $\zeta$, $\theta_{1}$, and
$\theta_{2}$, respectively.  

The effect of conical leads has already been studied in earlier
works (e.g., Ree and Reichl \cite{suhan1}, Persson et
al. \cite{conleads1}, and Berggren, Ji, and Lundberg
\cite{conleads2}). Compared to straight leads (which have most
commonly been used in publications on waveguides), conical leads are
more similar to the electron sea which provides the source of
electrons in most experiments. As shown in Ref.~\CITE{suhan1}, conical
leads allow tunneling resonances through the waveguide cavity for
energy regimes in which conduction is prohibited for straight leads.

A classical particle moving in a closed billiard whose shape is that
of a circle with a cut (the shape of our cavity) will display a rich
range of chaotic behavior as the size of the cut is varied
\cite{ergod}. Quantum signatures of the classical chaos in this
billiard have been studied in Ref.~\CITE{suhan2}. One question we will
be interested in is how well the scattering process can probe this
dynamics of the closed cavity. In the low-energy regime we will
consider, we expect that the waveguide will exhibit well-separated
Fano resonances at energies close to the energy eigenvalues of the
closed billiard.

Beyond simply allowing to probe eigenenergies of the closed billiard
as scattering resonances, lead placement itself can alter the symmetry
properties of the closed billiard. We will study the effect of
symmetries by investigating how the positions of the leads affect the
behavior of the Fano resonances.

In Sec.~2, we will discuss cavity states and the scattering states
that are used to obtain the scattering S-matrix for a system with
conical leads. In Sec.~3, quantum properties of the closed circular
cavity with a cut will be reviewed. In Secs.~4 and~5, we will look at
the behavior of conductance and Wigner delay times for waveguides with
full-circle and cut-circle cavities. Leads will be attached to
cavities symmetrically or asymmetrically. Finally, in Sec.~6, we will
summarize our results. A short discussion of the numerical method used
for our computations is given in the Appendix.

\section{Scattering wavefunction and S-matrix}
\label{sec:modes}
Matter waves inside the waveguide pictured in Fig.~\ref{fig:geo} are
governed by the Schr\"odinger equation, which, in polar coordinates,
takes the form
\begin{equation}
({\nabla}^2+k^2){\Psi}(\rho,\varphi)= {\biggl(}{{\partial}^2\over
{\partial}{\rho}^2}+{1\over {\rho}}{{\partial}\over
{\partial}{\rho}} +{1\over {\rho}^2}{{\partial}^2\over
{\partial}{\varphi}^2}
+{2mE\over {\hbar}^2}{\biggr)}{\Psi}(\rho,\varphi)=0.
\label{eq:schrod}
\end{equation}
Here, $k=\sqrt{2mE/{\hbar}^2}$ is the wavevector of the particle wave,
$m$ is the particle mass, $\hbar$ is Planck's constant, and $E$ is
the energy.  The geometry of the system, which determines its
scattering properties, is taken into account by the boundary condition
that the scattering wavefunction $\Psi$ vanish on the walls.

The solution $\Psi$ is computed numerically by expanding it into
suitable basis functions, and determining the expansion coefficients
from the Schr\"odinger equation and its boundary conditions by the
method described in the Appendix. Below, we will discuss the
expressions for $\Psi$ in the different regions of our system. In
Sec.~\ref{cavstates}, we look at the wavefunction inside the
cavity. The wavefunction inside the leads is considered in
Sec.~\ref{leadstates}, where we will also define the scattering
S-matrix of the waveguide.

\subsection{Cavity states}
\label{cavstates}
In the interior region of the cavity, the solution of
Eq.~(\ref{eq:schrod}) can be written as
\begin{equation}
\label{eq:cavstate}
\Psi ={\sum_{\gamma=-n_\gamma+1}^{n_\gamma}} c_\gamma f_{\gamma}\:.
\end{equation}
Here, the $f_{\gamma}$'s are suitable expansion functions. The
$c_\gamma$'s are the expansion coefficients, which must be determined
from the boundary conditions, by the procedure described in the
Appendix. The cutoff number $n_\gamma$ has to be chosen in a proper
way to achieve optimal numerical accuracy within a reasonable
computation time. We used $n_\gamma=90$ for all computations presented
here.

For the functions $f_\gamma$, several choices are possible in
principle. For our geometry, it proved to be numerically most
efficient to use solutions of the Helmholtz equation [$(\nabla^2+k^2)
f_\gamma=0$, without boundary conditions] which can be separated into
polar coordinates, ($\rho,\varphi$). These functions take the form
\begin{equation}
\label{=basis-cavity}
f_{\gamma}(\rho,\varphi)=J_{|\gamma|}(k\rho) \:e^{i \gamma \varphi}\:,
\end{equation}
where $J_{|\gamma|}$ denotes the Bessel function of order
${|\gamma|}$.

\subsection{Lead states}
\label{leadstates}

For waveguide scattering problems, it is usual to express the
properties of the system in terms of an S-matrix, which contains
reflection and transmission amplitudes from incoming channels to
outgoing channels in the leads. Before we can properly define the
S-matrix for our system, we first have to specify what we mean by
incoming and outgoing channels (also called ``modes'') in this
particular case.

Following the conventions from scattering problems with straight
leads, we require that an incoming (outgoing) mode $\chi^-$ ($\chi^+$)
is a solution of
\begin{equation}
\label{eq:helmholtz}
(\nabla^2+k^2) \chi^\pm =0
\end{equation}
which is separable into coordinates longitudinal and transversal to
the lead walls. As a further condition, the transverse part has to be
a standing wave vanishing on the walls and the longitudinal part
describes a wave propagating towards (away from) the cavity.

For the conical shape of our leads, the longitudinal and transversal
coordinates are just the polar coordinates $\rho$ and $\varphi$,
respectively. Thus, the separation ansatz with a standing wave in the
transverse direction ($\alpha=1,2,\ldots$) takes the form
\begin{equation}
{\chi}_\alpha^{i\pm}=Nu_{\alpha}^{i\pm}(\rho)~\sin \left[
{{\alpha}{\pi}{\left(\varphi-\theta_i+{\Delta\theta_i\over 2}\right)}
\over{\Delta}{\theta_i}} \right],
\label{eq:sep1}
\end{equation}
where $\theta_i$ locates the angular position of the center of the
$i$th lead, $R \leq \rho <\infty$, and $\theta_i-\Delta\theta_i /2
\leq \varphi < \theta_i+ \Delta\theta_i /2$ ($i=$1,~2 for leads~1
and~2, respectively). $N$ is a (yet unknown) constant
factor.

Inserting this into Eq.~(\ref{eq:helmholtz}) leads to an equation for
the radial part of the wavefunction,
\begin{equation} \label{=uequation}
\left[{\rho ^2}{{\partial}^2\over \partial{\rho^2}}+\rho{\partial
\over{\partial\rho}}-\left(
\frac{\alpha\pi} {\Delta\theta_i}\right)^2+k^2\rho^2 \right]
u^{i\pm}_\alpha(\rho)=0 \:.
\end{equation}
This is the Bessel differential equation and therefore
$u^{i\pm}_\alpha(\rho)$ can be expressed as a linear combination of
Bessel functions, $J_{\alpha\pi/\Delta\theta_i}(k\rho)$ and
$Y_{\alpha\pi/\Delta\theta_i}(k\rho)$. Since we required that the
longitudinal part represent a propagating wave, we write the radial
solution in terms of Hankel functions,
\begin{equation} \label{=ansatzu-cl}
u_\alpha ^\pm (\rho )= H^\pm_\frac{\alpha\pi}{\Delta\theta_i} (k\rho)=
J_\frac{\alpha\pi}{\Delta\theta_i}(k\rho ) \pm
iY_\frac{\alpha\pi}{\Delta\theta_i}(k\rho )\:,
\end{equation}
which approach exponential functions at infinity:
\begin{equation} \label{=asymptotic2}
H^\pm_\frac{\alpha\pi}{\Delta\theta_i}(k\rho )
\stackrel{k\rho\rightarrow\infty}{\rightarrow} 
\sqrt{\frac{2}{\pi k\rho}} \: \exp \left\{ 
\pm i\left[k\rho -\frac{\pi}{4}\left(2\frac{\alpha\pi}
{\Delta\theta_i}+1\right)\right] \right\} 
+{\cal O} \left[ \frac{1}{(k\rho)^{3/2}}\right] \:.
\end{equation}
Here, we used the notation $H_\gamma^+$ ($H_\gamma^-$) for the Hankel
functions of first (second) kind, of order $\gamma$.

The normalization constant, $|N|$, is calculated from the condition
that each channel carry unit current, i.e., we require that
\begin{equation}
\label{eq:unitcur}
\left| \int_{\theta_i-\Delta\theta_i/2} ^ {\theta_i+\Delta\theta_i/2} 
\rho\: d\varphi \: j^{i\pm}_\rho\right| = 1.
\end{equation}
Here, the longitudinal component of the probability current density,
${\bf j}^{\,i\pm} = j^{\,i\pm}_\rho \: {\bf e_\rho}+
j^{\,i\pm}_\varphi \: {\bf e_\varphi}$, is given by
\begin{equation} 
\label{=current}
j^{\,i\pm}_\rho=\frac{i\hbar}{2m} \left[ 
\left( \frac{\partial \chi^{i\pm}} {\partial \rho} \right)^*
\chi^{i\pm} - 
\chi^{i\pm^*}\left( \frac{\partial \chi^{i\pm}}
{\partial \rho} \right) \right].
\end{equation}
This yields
\begin{equation} 
\label{=normal-cl-abs}
|N|=\sqrt{\frac{\pi m}{\hbar\Delta\theta_i}}\:.
\end{equation}

With these results, the expression for the incoming/outgoing channel
$\alpha=1,2,\ldots$, corresponding to the $\alpha$th standing wave in
transverse direction, becomes
\begin{equation} 
\label{=mode-cl-abs}
\chi_\alpha^\pm = \frac{N}{|N|}
\sqrt{ \frac{m\pi} {\hbar\Delta\theta_i} }\; 
\sin \left[ \frac{\alpha \pi 
{\left(\varphi - \theta_i + \frac{\Delta\theta_i}{2} \right)}}
{\Delta\theta_i} \right]\:
H^\pm_\frac{\alpha\pi}{\Delta\theta_i}(k\rho)\:.
\end{equation}

To fix the overall phase factor, $N/|N|$, in Eq.~(\ref{=mode-cl-abs}),
we require that the solution be real at the interface between the
leads and the cavity. Then, if a mode is completely reflected back at
this position, the corresponding reflection coefficient will be 1.
This can be achieved by multiplying the modes with a complex phase
factor. At $\rho=R$, the Hankel functions (the only complex factor in
$\chi_\alpha^{i\pm}$) can be written as
\begin{equation} \label{=hankel-phase}
H_{\frac{\alpha\pi}{\Delta\theta_i}}^\pm (kR)=
\left| H_{\frac{\alpha\pi}{\Delta\theta_i}}^\pm (kR) \right| \:
\exp\left\{ \pm i 
\arctan \left[\frac{Y_{{\alpha\pi}/{\Delta\theta_i}}(kR)}
{J_{{\alpha\pi}/{\Delta\theta_i}}(kR)}\right] \right\} \:.
\end{equation}
Because we want the probability amplitude at this position to be real, we
will use the complex phase factor
\begin{equation} \label{=normal-cl}
\frac{N}{|N|}= \exp \left\{ \mp i \:
\arctan \left[ \frac{Y_{{\alpha\pi}/{\Delta\theta_i}}(kR)}
{J_{{\alpha\pi}/{\Delta\theta_i}}(kR)}\right] \right\} \:.
\end{equation}

Then, the final form of the incoming and outgoing propagating modes in
channel $\alpha$ becomes
\begin{equation} 
\label{eq:mode}
\chi_\alpha^{i\pm} (\rho,\varphi)=
\sqrt{\frac{m\pi}{\hbar\Delta\theta_i}}\;
\exp \left\{ \mp i \:
\arctan \left[\frac{Y_{\frac{\alpha\pi}{\Delta\theta_i}}(kR)}
{J_{\frac{\alpha\pi}{\Delta\theta_i}}(kR)}\right] \right\}
\sin \left[ \frac{\alpha \pi {\left(\varphi-\theta_i+
\frac{\Delta\theta_i}{2}\right)}}{\Delta\theta_i} \right]\:
H^\pm_\frac{\alpha\pi}{\Delta\theta_i} (k\rho)\:.
\end{equation}

As we now have derived the expression for incoming and outgoing
channels, we are finally in the position to define the S-matrix of our
system. The matter wave $\Psi$ inside the leads is expanded in terms
of the incoming and outgoing modes, $\chi_{\alpha}^{i-}
(\rho,\varphi)$ and $\chi_{\alpha}^{i+} (\rho,\varphi)$,
respectively. Then the expansion coefficients describe the probability
amplitudes of reflection or transmission of the matter wave from an
incoming channel to the outgoing channels. More precisely, if the
electron is incident in lead 1 and channel $\beta$, the wavefunction
in the leads takes the form
\begin{equation}
\label{eq:leadexp1}
\Psi_{\beta,1} =
\chi^{1-}_\beta+\sum_\alpha
r_{\alpha\beta}\chi^{1+}_\alpha 
~~~{\rm and}~~~\Psi_{\beta,2} = \sum_\alpha
t_{\alpha\beta}\chi^{2+}_\alpha\:.
\end{equation}
Here, $\Psi_{\beta,1}$ is the state in lead 1, $\Psi_{\beta,2}$ is the
state in lead 2, and $r_{\alpha\beta}~(t_{\alpha\beta})$ is the
probability amplitude that the electron enters in channel $\beta$ and
is reflected (transmitted) into channel $\alpha$.  Likewise, if the
electron is incident in lead 2 and channel $\beta$, the wavefunction
in the leads takes the form
\begin{equation}
\label{eq:leadexp2}
{\Psi}_{\beta,2} =
\chi^{2-}_\beta+\sum_\alpha
r^{'}_{\alpha\beta}\chi^{2+}_\alpha 
~~~{\rm and}~~~\Psi_{\beta,1} =
\sum_\alpha t^{'}_{\alpha\beta}\chi^{1+}_\alpha\:.
\end{equation}
Here $\Psi_{\beta,2}$ is the state in lead 2, $\Psi_{\beta,1} $ is the
state in lead 1, and $r^{'}_{\alpha\beta}$ and $t^{'}_{\alpha\beta}$
are the reflection and transmission probability amplitudes,
respectively. Note that the solution $\Psi$ of Eq.~(\ref{eq:schrod})
is not unique, but we obtain different solutions $\Psi_\beta$
depending on which channel $\beta$ is assumed to carry the incoming
part of the matter wave.

The S-matrix connects the amplitudes of the incoming propagating
electrons to the amplitudes of the outgoing propagating electrons and
it is therefore constructed out of all the reflection and transmission
amplitudes. Thus, the S-matrix can be written
\begin{equation}
{\bf S}=\left( \begin{array}{cc} {\bf R}&{\bf T'}\\{\bf
T}&{\bf R'}\end{array} \right) \:,
\end{equation}
where $\bf R$ and $\bf R'$ are square matrices composed of reflection
amplitudes, $r_{\alpha\beta}$ and $r'_{\alpha\beta}$, and $\bf T$ and
$\bf T'$ are square matrices composed of transmission amplitudes,
$t_{\alpha\beta}$ and $t'_{\alpha\beta}$. The reflection and
transmission amplitudes must be computed numerically (see Appendix).

\section{Closed cut-circle billiard}
\label{sec:circlecut}
One aspect of the scattering properties of the waveguide that we wish
to explore is how close a correspondence there is between the Fano
scattering resonances and the eigenstates of a closed version of the
waveguide cavity. In this section, we therefore describe some
features of these eigenstates as the cut size on the cut-circle
billiard is changed.

The Schr\"odinger equation [Eq.~(\ref{eq:schrod})] for a quantum
particle in a closed full-circle billiard is separable, with one part
describing the radial and the other part the azimuthal motion.  The
(discrete) energy spectrum can be labeled with two good quantum
numbers.  One of them, $n=1,2,...$, is associated with radial motion,
and the other, $l=0,{\pm}1,{\pm}2,...$, with the angular motion.  The
eigenfunctions are
\begin{equation}
\Psi_{l,n}=J_{|l|} \left( k_{|l|,n}~\rho \right) e^{il\varphi} \:, 
\label{circle-ef}
\end{equation}
where $J_{|l|}$ is the Bessel function of order $|l|$ and $(k_{|l|,n}
R)$ is its $n$th zero (i.e., $J_{|l|}$ vanishes at $\rho=R$). The
corresponding eigenenergies $E_{|l|,n}$ are then calculated via
$E=\hbar^2 k^2/ (2m)$. The energy eigenstates of the circle billiard
with angular momentum $|l|{\geq}1$, corresponding to clockwise or
counter clockwise rotation of the matter wave, produce two-fold
degenerate eigenenergies. Only the eigenenergies with $l=0$ are
non-degenerate.

If we apply a cut to the closed system, the classical dynamics of the
particle exhibits hard chaos \cite{ergod}, as long as
${\omega}{\leq}180^\circ$. Looking at the quantum properties, we
observe a splitting of the degenerate eigenenergies as the cut is
inserted. The absence of degenerate eigenvalues (level repulsion) is a
well-known quantum feature of classically chaotic systems (see, e.g.,
Ref.~\CITE{todd}).

\begin{figure}[h]
\begin{center}
\includegraphics[scale=0.7]{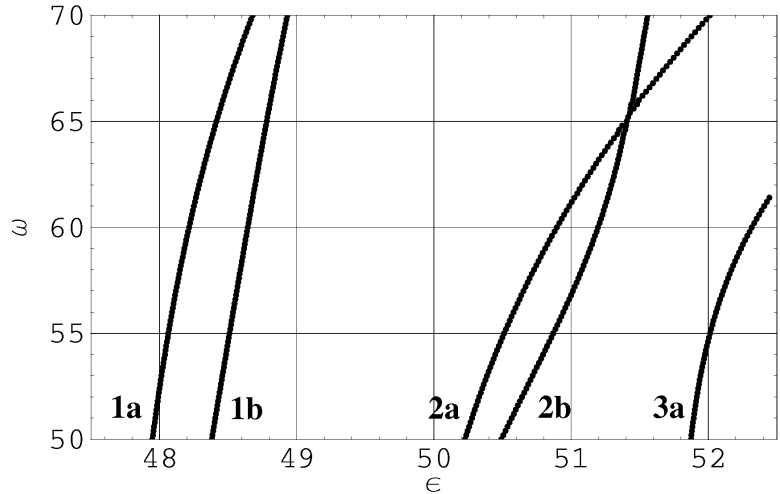}
\caption {Five energy eigenvalues of the closed cut billiard as a
function of dimensionless energy $\epsilon$ and cut size
$\omega$. From left to right we label them as 1a, 1b, 2a, 2b, and
3a. For $\omega=0^\circ$, the states 1a and 1b are degenerate with
quantum numbers $(l=3,n=2)$, the states 2a and 2b are degenerate with
$(l=6,n=1)$, and 3a is one of a degenerate pair with $(l=1,n=3)$.
\label{fig:bileigval}}
\end{center}
\end{figure}

Examining the quantum properties of the cut-circle billiard in more
detail, we show how five of the energy levels of the cut billiard
behave as the cut size is varied from ${\omega}=50^\circ$ to
${\omega}=70^\circ$ in Fig.~\ref{fig:bileigval}. The states shown are
continuations of the circle cavity degenerate pairs $(l=3,n=2)$,
$(l=6,n=1)$, and one of the pair, $(l=1,n=3)$.

Here and in later numerical computations, we are using dimensionless
variables. The dimensionless energy $\epsilon$ is given by
\begin{equation}
{\epsilon}={mR^2\over {\hbar}^2}E={(kR)^2\over 2}={{\kappa}^2\over 2},
\label{dim-en}
\end{equation}
where ${\kappa}=kR$ is a dimensionless wavevector. To get an idea of
how this translates to the physical dimensions of our system, consider
an electron density of $n \:{\approx}\: 4{\times} 10^{11}\: {\rm
cm}^{-2}$, which is a typical value measured for a two dimensional
electron gas in a GaAs/Al$_{0.3}$Ga$_{0.7}$As heterostructure
\cite{marcus}. The corresponding Fermi wavevector is $k_f =
{\sqrt{2\pi n}} \:{\approx}\: 1.6{\times}10^8 \:{\rm m}^{-1}$, the
Fermi energy $E_f = \pi\hbar^2 n/m \approx 0.014 \:{\rm eV}$ (for an
effective electron mass in GaAs of $m=0.067 m_e$). This yields for
$\epsilon=100$, which is the maximum energy we used in the
computations, $R \:\approx\: 90 \:{\rm nm}$.

In Fig.~\ref{fig:bileigval}, we observe a (distant) avoided crossing,
which occurs between states 2b and 3a in the interval
$\omega=53^\circ$ to $\omega=63^\circ$. We also see an actual
crossing in this plot. This occurs between the states labeled 2a and
2b at about $\omega=65^\circ$.

However, as discussed in Ref.~\CITE{wigner}, separate eigenvalues of a
generic Hamiltonian can in general only be brought to coincide if at
least three parameters are varied. Thus, at first sight, we would
expect another level repulsion instead of the actual crossing, since
we only vary $\omega$. Yet we ignored that the Hamiltonian of the
cut-circle is not completely ``generic''. Even if the cut is inserted,
the system retains parity symmetry with respect to an axis
perpendicular to the cut, i.e.,
\begin{equation}
\label{eq:parity}
P_\zeta \Psi(\rho,\varphi) \equiv
\Psi(\rho, \zeta-\varphi) = \Psi(\rho,\varphi) \:,
\end{equation}
where $P_\zeta$ is the parity operator for a symmetry axis through the
center of the cavity, at angle $\zeta$ (i.e., the angle at which the
center of the cut is placed). Because of parity symmetry, we have two
different symmetry classes, namely the classes formed by even and
odd states. Since parity is a discrete symmetry, states of one class
cannot interact with states of the other, thus both classes are
completely independent of each other. Therefore, crossings between
states of different parity may occur. Only {\em within} each parity
class, i.e., among states with the same parity, crossings remain
impossible.

\begin{figure}[h] 
\begin{center} 
\includegraphics[scale=1.0]{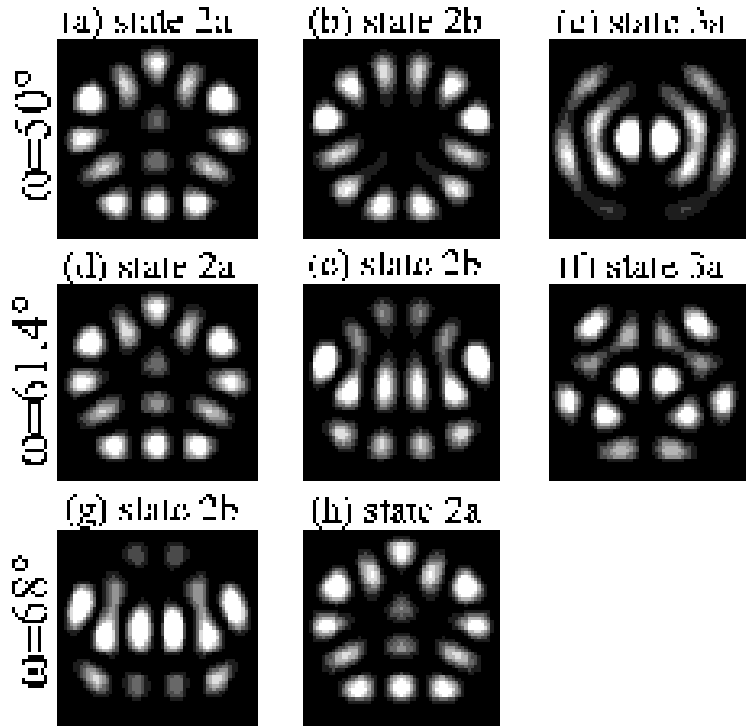}
\caption {Three energy eigenstates corresponding to energy eigenvalues
2a, 2b, and 3a in Fig.~\ref{fig:bileigval}. The cut is located at
$\zeta=270^\circ$. The top row has $\omega=50^\circ$ and consists of
(a) State 2a for $\epsilon=50.2$, (b) State 2b for $\epsilon=50.5$,
and (c) State 3a for $\epsilon=51.9$. The middle row has
${\omega}=61.4^\circ$ and consists of (d) State 2a for
$\epsilon=51.0$, (e) State 2b for $\epsilon=51.3$, and (f) State 3a
for $\epsilon=52.4$. The bottom row has $\omega=68^\circ$ and consists
of (g) State 2b for $\epsilon=51.5$ and (h) State 2a for
$\epsilon=51.7$. Note that, from left to right, we show states with
increasing energy. This means that 2a and 2b switch position from the
middle to the bottom row, as their eigenenergies cross.
\label{fig:bileigstate}} 
\end{center} 
\end{figure}

In Fig.~\ref{fig:bileigstate}, we show energy eigenstates
corresponding to the eigenvalues 2a, 2b, and 3a in
Fig.~\ref{fig:bileigval}, for three different cut sizes taken before
the avoided crossing, between the avoided and actual crossings, and
after the actual crossing. It can be seen that 2a has even parity, and
2b and 3a both have odd parities. This explains why 2a and 2b can
undergo the crossing, whereas a crossing between 2b and 3a is avoided.

We also see that the state 2a, which does not participate in the
avoided crossing and undergoes the crossing with 2b, remains
unchanged.  However, states 2b and 3a, which do undergo an avoided
crossing, become mixed and loose their original identities. The mutual
parity of states 2b and 3a allows them to couple, whereas this is
impossible for states 2a and 2b with differing parities.

\section{Waveguide with a circle cavity}
\label{sec:circguide}

If we attach leads to the circle billiard, we have a matter waveguide
and the eigenstates of the billiard become unstable and may decay.
These unstable states strongly affect the scattering dynamics of the
waveguide. They appear as resonances in the transmission probabilities
through the waveguide,
\begin{equation} 
\label{=ti}
T_\beta=\sum_{\alpha} |t_{\alpha\beta}|^2 \:,
\end{equation}
for incoming mode $\beta$ in lead~1, where $\alpha$ runs over the
outgoing modes in lead~2. We can examine the individual $T_\beta$'s or
consider their sum over all possible incoming modes in lead~1, which
is proportional to the Landauer-B\"uttiker conductance
\cite{landauer},
\begin{equation}
\label{=G}
G =\frac{2e^2}{h}\sum_\beta T_\beta \:.
\end{equation}

As shown in Ref.~\CITE{suhan1}, there is an energy range where, for
conical leads, mode $\beta$ has nonzero transmission, whereas for
straight leads with the same opening angles it would still be
evanescent. This occurs for energies too small for the $\beta$th
transverse standing wave to fit into the lead opening at the junction
to the cavity, such that the wave has to ``tunnel'' into the
cavity. The condition for this tunneling (see Ref.~\CITE{suhan1}) is
$k< \beta\pi/d_i$, if $d_i=2R \sin (\Delta\theta_i/2)$ is the width of
the lead opening (for opening angle $\Delta\theta_i$). Thus, tunneling
occurs for energies such that
\begin{equation} \label{=tunnel}
\epsilon < \tilde\epsilon_\beta=\frac{\beta^2 \pi^2 R^2}{2 d_i^2}=
\frac{\beta^2 \pi^2}{8\sin^2(\Delta\theta_i/2)} \:,
\end{equation}
where $\epsilon$ is the dimensionless energy defined in
Eq.~(\ref{dim-en}).

In addition to the transmission probabilities and Landauer-B\"uttiker
conductance, it is useful to look at the Wigner delay times,
\begin{equation} 
\label{=wigner-td}
\tau_n^{W}=\hbar \frac{\partial \theta_n}{\partial E}\:,
\end{equation}
where $\theta_n$ are the phases of the S-matrix eigenvalues
$s_n=e^{i\theta_n}$. The Wigner delay times are a measure of the time
delay of the electron due to the presence of the cavity. In the
following plots, we will use dimensionless delay times
\begin{equation}
\label{=td-conversion}
\tau_n=\frac{\partial \theta_n}{\partial \epsilon}
=\frac{\hbar\tau^W_n}{m_e^*R^2} \:.
\end{equation}
(For $R\approx 100\:{\rm nm}$, $\tau_n \approx 20$, and $m_e^*=0.067
m_e$, the Wigner delay time is of the order of 0.1 ns.) From
Ref.~\CITE{na2} it is clear that peaks in the Wigner delay time
correspond to resonances and poles of the scattering matrix in the
complex energy plane. Therefore we can use peaks in the delay time
spectra to indicate the resonances of our system.

\subsection{Symmetric placement of leads}
\label{circguidesym}

\begin{figure}[h]
\begin{center}
\includegraphics[scale=0.7]{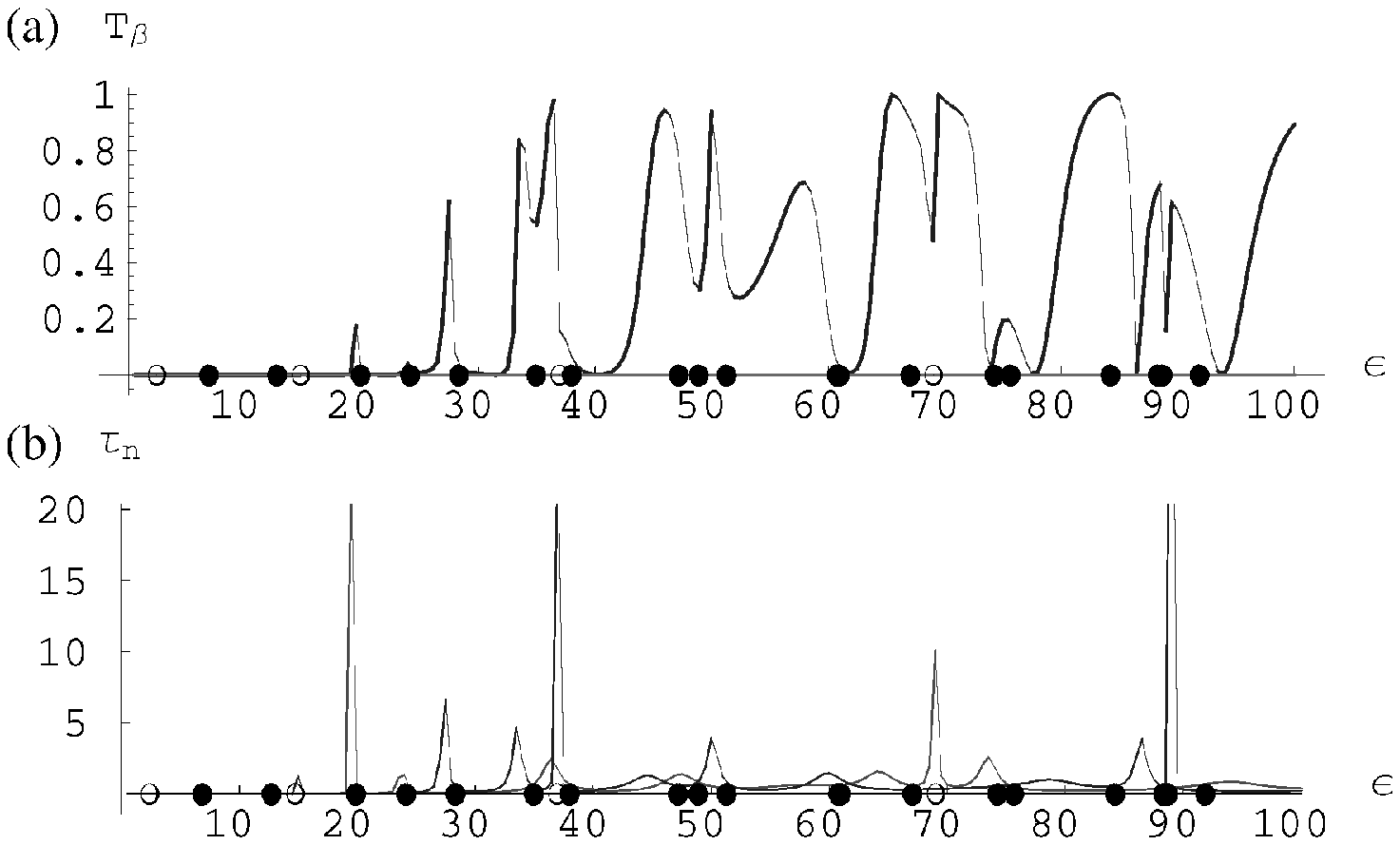}
\caption {(a) Transmission probability and (b) delay time spectra for
lead openings $\Delta\theta_1=20^\circ$, $\Delta\theta_2=20^\circ$ and
lead positions $\theta_1=0^\circ$ and $\theta_2=180^\circ$.
Eigenvalues of the closed circle are marked by filled (degenerate
eigenvalues) and open (non-degenerate eigenvalues) circles.
\label{fig:wgcirc}}
\end{center}
\end{figure}

In Fig.~\ref{fig:wgcirc}, we show numerical data obtained for an
electron waveguide with a circle cavity with symmetrically placed
leads, one at $\theta_1=0^\circ$ and the other at
$\theta_2=180^\circ$. Both leads subtend the same angle
$\Delta\theta_1 = \Delta\theta_2 = 20^\circ$. The dimensionless energy
$\epsilon$ is varied in steps of 0.5.


The transmission probability in Fig.~\ref{fig:wgcirc}(a) shows the
quantities $T_\beta$ for incoming modes $\beta=1,2$ in lead~1. The
first mode ($\beta=1$) starts transmitting (i.e., becomes
significantly larger than zero) at about $\epsilon \approx 20$. The
second mode starts transmitting too closely below 100 to be visible in
the plot. The transmission for higher modes is zero everywhere in the
energy interval below 100, and therefore these modes were not included
in the computations or the plot.

The threshold energies $\tilde\epsilon_\beta$, for which the modes
would start propagating in the case of straight leads with
$\Delta\theta=20^\circ$, are $\tilde\epsilon_1=40.91$, and
$\tilde\epsilon_2=163.66$ for the first two channels. This means that
for the conical leads, we observe tunneling of the first mode in the
energy range between $\epsilon \approx 20$ and 40.

Examining the correspondence between eigenenergies of the closed
circle (indicated by filled circles for degenerate levels and open
circles for non-degenerate levels) and resonances of the open system,
one finds that there is obviously a strong association. In energy
intervals without eigenenergies, the transmission changes smoothly,
whereas there are sharp changes close to eigenenergies. In the
tunneling regime, we see peaks close to the eigenenergies. The peaks
tend to be slightly shifted to the left. Above the tunneling
threshold, we also observe valleys, e.g., at $\epsilon=67.5,74.7$. In
this regime, most peaks and valleys in the vicinity of an eigenenergy
are too close together to tell which is associated with the
eigenenergy. Since the lead openings themselves alter the geometry of
the system, we cannot expect the resonances to be found exactly at the
same positions as the eigenenergies.

In Fig.~\ref{fig:wgcirc}(b), we plot the dimensionless delay times
$\tau_n$. The energy derivatives at each point are approximated by
carrying out two computations for each point, with energy spacing
$\Delta \epsilon=10^{-5}$. Since the transmission probabilities
$T_\beta$ vanish for $\beta\geq 3$, we only have to consider the two
channels $\beta=1$ and $\beta=2$. Thus, the S-matrix is a 4x4 matrix
and we get four eigenphases and four delay time curves.

Here, unlike for the transmission probabilities, we only need to look
for {\em peaks} in the Wigner delay times at the resonance energies,
in the tunneling as well as in the normal regime. The first two time
delays display their first peak around $\epsilon\approx 15$, the
others at $\epsilon \approx 75$. At these energies, the corresponding
transmission probabilities are still too close to zero to produce a
visible peak in Fig.~\ref{fig:wgcirc}(a). This shows that the delay
times are a more sensitive indicator of resonances than the
$T_\beta$'s.

We see that the degenerate eigenenergies of the circle billiard do not
produce double resonances when leads are attached at the special
angles $\theta_1=0^\circ$ and $\theta_2=180^\circ$. Every eigenenergy
of the circle billiard is associated only with a {\em single} delay
time peak. Although placing leads to the circle cavity destroys
its radial symmetry, the breaking of the degeneracy is not observed in
this case.

%
%
\subsection{Asymmetric placement of leads}
\label{circguideasym}

\begin{figure}[h]
\begin{center}
\includegraphics[scale=0.7]{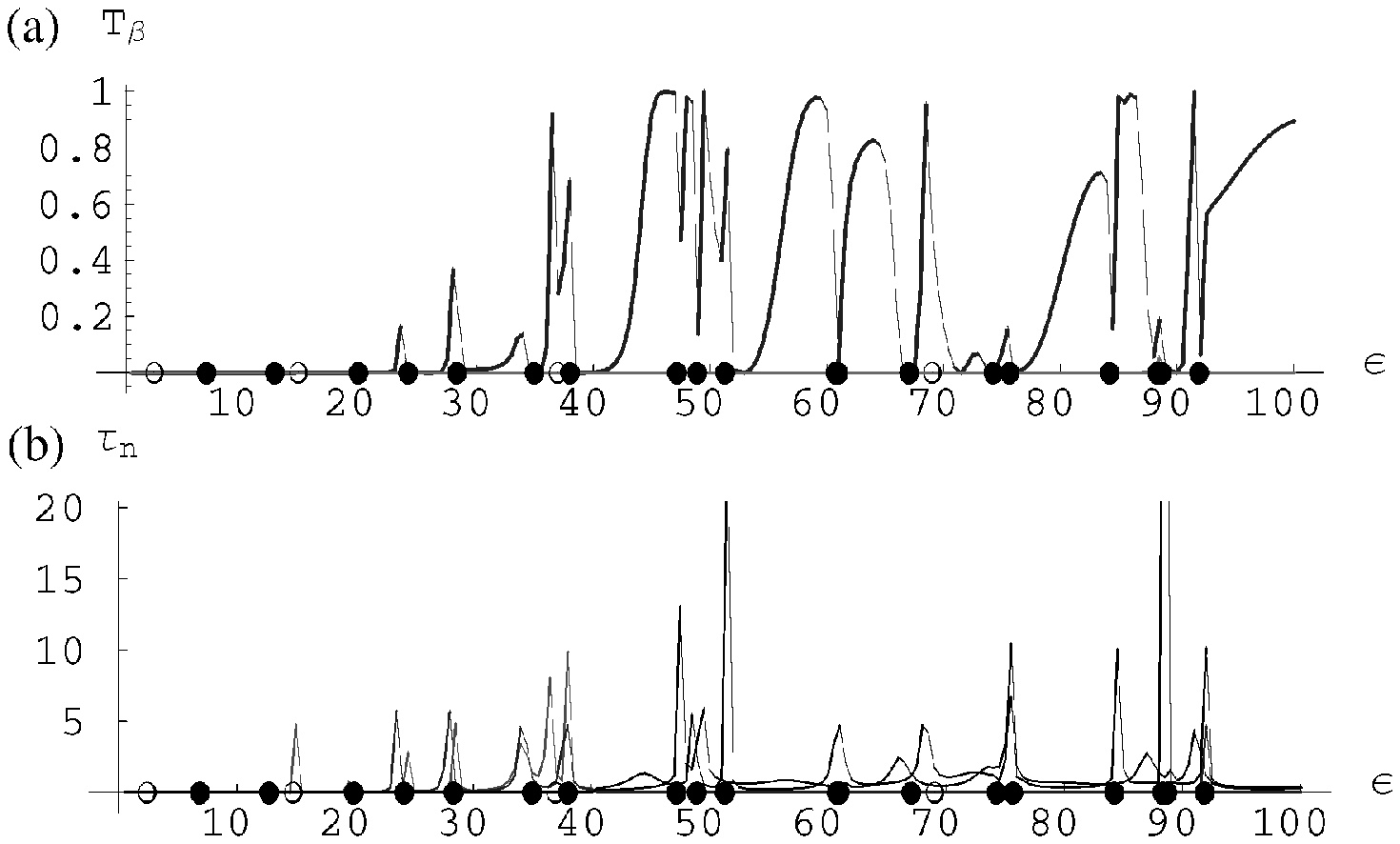}
\caption {(a) Transmission probability and (b) delay time spectra for
lead openings $\Delta\theta_1=20^\circ$, $\Delta\theta_2=20^\circ$ and
lead positions $\theta_1=0^\circ$ and $\theta_2=125^\circ$.
Eigenvalues of the closed circle are marked by filled (degenerate
eigenvalues) and open (non-degenerate eigenvalues) circles.
\label{fig:wgasymcir}}
\end{center}
\end{figure}

In Fig.~\ref{fig:wgasymcir}, we show the same waveguide, but with
asymmetric placement of leads instead of the symmetric placement we
had in Fig.~\ref{fig:wgcirc}. Lead~1 is placed at $\theta_1=0^\circ$
and lead~2 at $\theta_2=125^\circ$.


Now we not only see again the close correspondence between delay time
peaks and eigenenergies of the circle, but we also observe double
peaks at most of the degenerate billiard eigenenergies, most evidently
at $\epsilon=$24.6, 28.8, 35.4, 38.5, and 49.4. At higher energies,
the observation of double peaks is more difficult because the second
mode starts transmitting. (Some of its first peaks are too small to be
seen in the plot, but can be found by looking at the data file.) Many
of the double peaks show a small energy spacing: One peak is slightly
shifted to smaller energies, the other one to higher energies. Here,
the degeneracy is obviously broken by the addition of the leads. For
the non-degenerate eigenenergies on the other hand, no splitting can
be observed. Especially at $\epsilon=37.4$, this becomes very obvious.

The difference between this case and the previous case with
symmetric lead placement ($\theta_2=180^\circ$) is that, for symmetric
placement, we had two discrete symmetries, namely invariance under the
parity transformations $P_{0^\circ}$ and $P_{90^\circ}$ [where the
parity operators are defined as in Eq.~(\ref{eq:parity})]. When the
leads are placed asymmetrically, the symmetry under $P_{0^\circ}$,
with respect to the axis bisecting both leads, is destroyed, and we
are only left with one parity symmetry under $P_{62.5^\circ}$. The
breaking of one parity symmetry seems to be the reason for the
splitting of resonances seen in Fig.~\ref{fig:wgasymcir}(b).

\subsection{Numerical accuracy}
\label{numacc}

To get an estimate of the numerical error, the sum of the absolute
values squared of all reflection and transmission amplitudes was
calculated. Due to the unitarity of the S-matrix, this value has to be
the dimension of the S-matrix, which is 4 here. The relative error,
\begin{equation} \label{=error}
1 - \frac{1}{4}\sum_{\alpha,\beta=1}^2
(|r_{\alpha\beta}|^2 + |t_{\alpha\beta}|^2 +
|r^{'}_{\alpha\beta}|^2 + |t^{'}_{\alpha\beta}|^2) \:,
\end{equation}
was usually smaller than 0.2\% for the individual data points. The
errors of the absolute values of the S-matrix eigenvalues (which
should be 1) were of the same order of magnitude. For the delay
times, there is no easy way to estimate the error.


The most crucial factor determining numerical accuracy was found to be
the number $2n_\gamma$ of parameters (expansion coefficients
$r_{\alpha\beta}, t_{\alpha\beta}, b_{\nu\beta}$) used to approximate
the total wavefunction $\Psi$ of our system [see
Eqs.~(\ref{eq:leadexp1}), (\ref{eq:wallexp}),
and~(\ref{=cutoffs})]. An estimate for a suitable choice of this
number can be found by comparing the Fermi wavelength, $\lambda_f =
2\pi /k_f$, to a typical length scale of the system, e.g., the length
of the boundary $\approx 2\pi R$ (for zero or small cut size). If we
assume that the total wavefunction of our system varies rather
smoothly on a length scale of $\lambda_f$, it seems plausible that a
``small'' number of parameters, say about 10, should be sufficient to
approximate $\Psi$ on an interval of length $\lambda_f$. Therefore, we
expect that such a ``small'' multiple of $\kappa=k_f R =
\sqrt{2\epsilon}$=``the number of $\lambda_f$-intervals needed to
cover the cavity boundary'' should give a reasonable order of
magnitude estimate for $2n_\gamma$.

\begin{figure}[h]
\begin{center}
\includegraphics[scale=1.0]{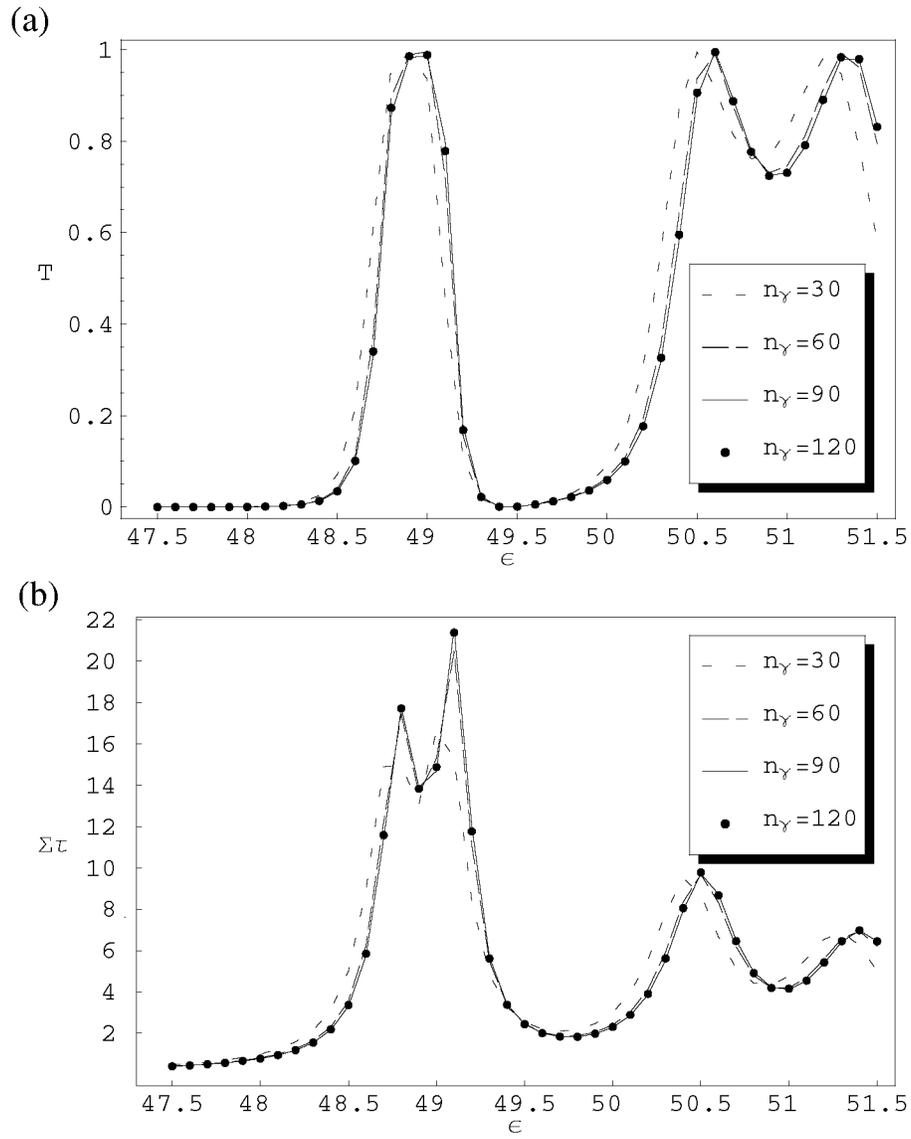}
\caption{(a) Transmission spectra and (b) spectra of the sum of
partial delay times in the dimensionless energy interval
$47.5{\leq}\epsilon{\leq}52.5$, for cut size
$\omega{\leq}80^\circ$. The lead openings are
$\Delta\theta_1=\Delta\theta_2=16^\circ$. Lead and cut positions are
$\theta_1=0^\circ$, $\theta_2=120^\circ$, and $\zeta=240^\circ$. The
results for different values of $n_\gamma$ are compared.
\label{fig:comp}}
\end{center}
\end{figure}

In Fig.~\ref{fig:comp}, we compare the transmission spectra
[Fig.~\ref{fig:comp}(a)] and the spectra of the time delay sums
[Fig.~\ref{fig:comp}(b)] for different values of $n_\gamma$. As
deviations are expected to be most pronounced for large cut size, the
calculations shown were done exemplarily for the symmetric cut-circle
waveguide to be examined in Sec.~\ref{cgsym}, at $\omega=80^\circ$.

We find noticeable deviations for $n_\gamma=30$ and, to a small
extent, also for $n_\gamma=60$. The values for $n_\gamma=90$
practically coincide with those for $n_\gamma=120$. Therefore, in view
of computation time, we used $n_\gamma=90$ in all computations
presented in this paper.

\section{Waveguide with a cut-circle cavity}
\label{sec:cutcircguide}
%
%

\begin{figure}[h]
\begin{center}
\includegraphics[scale=1.0]{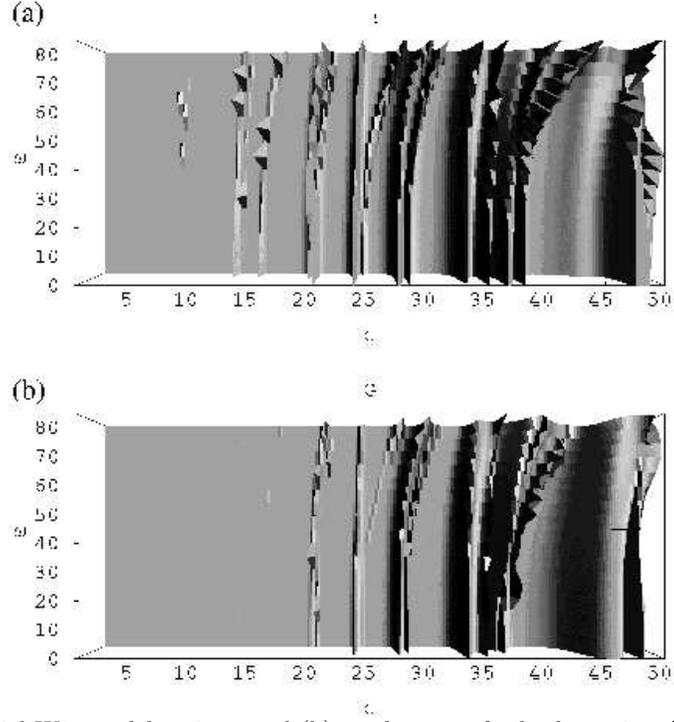}
\caption{(a) Sum of the partial Wigner delay times and (b) conductance
for lead opening $\Delta\theta=20^\circ$ and varying cut size
$\omega$.  The leads are centered at $\theta_1=0^\circ$ and
$\theta_2=166^\circ$, and the cut is centered at $\zeta=225^\circ$.
\label{fig:cut20}}
\end{center}
\end{figure}

We now consider a waveguide with a cut-circle cavity.
Fig.~\ref{fig:cut20} shows the sum of the partial delay times,
$\tau=\sum_ {n=1}^4\tau_n$, and the Landauer-B\"uttiker conductance
$G$. For this computation, leads and cut are placed
asymmetrically. The leads are centered at $\theta_1=0^\circ$ and
$\theta_2=166^\circ$, and the cut at $\zeta=225^\circ$. The cut size
$\omega$ varies from $0^\circ$ (circle) to $80^\circ$, with step size
$5^\circ$. The dimensionless energy $\epsilon$ is varied in steps of
0.25.

In the delay time spectrum (upper plot), we observe chains of
separated double peaks, corresponding to the split-up branches of
degenerate circle eigenenergies. If we compare the positions of maxima
in the Wigner delay time to the positions of eigenenergies of the
closed system, we find a good agreement between (open system)
resonances and (closed cavity) eigenenergies.

The conductance (lower plot) shows similar features. However, the
first visible chains of resonances occur at significantly higher
energies than in the delay time plot. For energies between 20 and 40,
both plots show basically the same structures. Above the first
tunneling threshold at $\tilde\epsilon_1=40.91$, the conductance
reflects the onset of ``normal'' (compared to tunneling) transmission
of the first mode. In this energy regime, we cannot tell any more
whether resonances are indicated by peaks or dips in the conductance.

We will now compare the properties of the cut-circle cavity
waveguide for symmetric and asymmetric lead placements. Note, however,
that although this may seem to be analogous to the analysis from the
previous section (for the full-circle cavity), the insertion of the
cut already leaves us with only one parity symmetry (under
$P_\zeta$). As a result, the ``symmetric'' cut-circle waveguide is
analogous to the {\em asymmetric} full-circle waveguide rather than to
the symmetric full-circle waveguide. Finally, in the ``asymmetric''
cut-circle waveguide, there are no geometric symmetries present at
all. Since this could not be achieved by merely placing two leads to a
circle, this case has no parallel in Sec.~\ref{sec:circguide}.

\subsection{Symmetric placement of leads and cut}
\label{cgsym}

\begin{figure}[h]
\begin{center}
\includegraphics[scale=1.0]{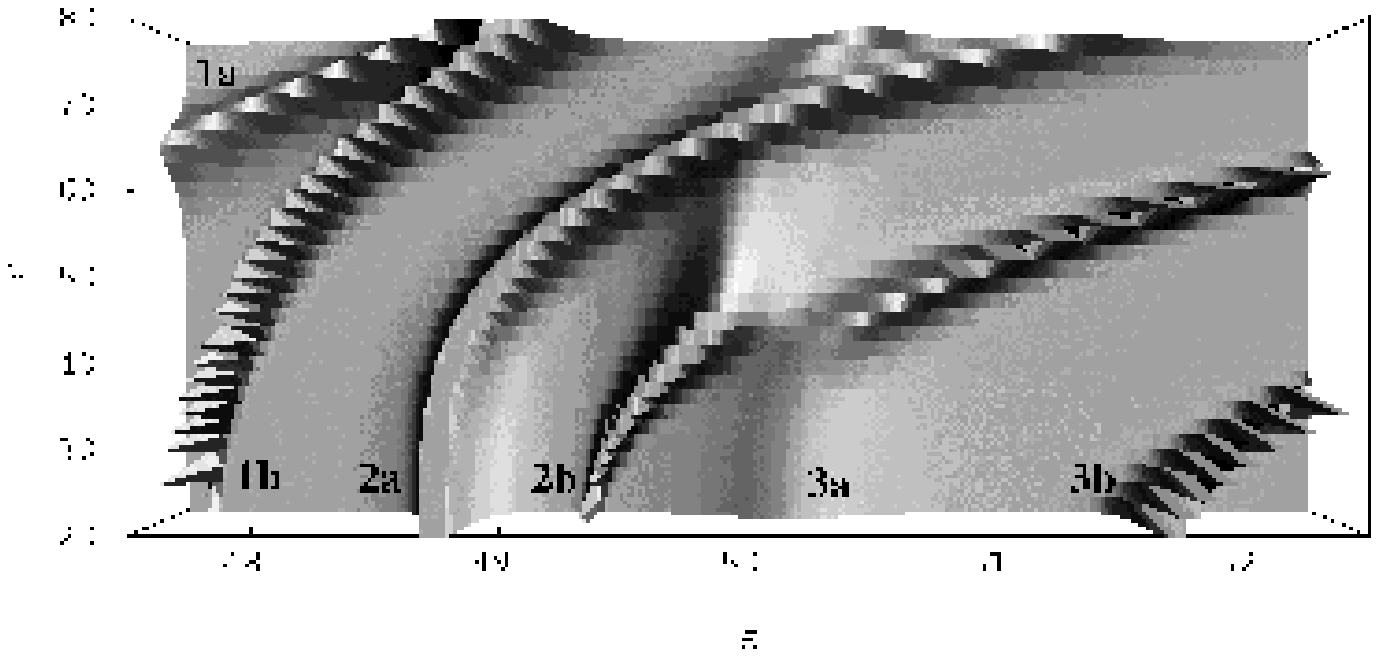}
\caption{Spectrum of the sum of partial delay times for the energy
interval $47.5{\leq}\epsilon{\leq}52.5$ and cut size interval
$20^\circ{\leq}\omega{\leq}80^\circ$. The lead openings are
$\Delta\theta_1=\Delta\theta_2=16^\circ$. Lead and cut positions are
$\theta_1=0^\circ$, $\theta_2=120^\circ$, and $\zeta=240^\circ$.
\label{fig:wgsymcut16}}
\end{center}
\end{figure}

Let us examine the Wigner delay times for the energy range where we
found an actual crossing and an avoided crossing in
Fig.~\ref{fig:bileigval} for the closed cut-circle billiard. We first
consider a system in which the leads are placed at $\theta_1=0^\circ$
and $\theta_2=120^\circ$, symmetrically with respect to the cut at
$\zeta=240^\circ$. The lead openings are $\Delta\theta_1 =
\Delta\theta_2 = 16^\circ$. The plot of the sum $\tau$ of the Wigner
delay times is shown in Fig.~\ref{fig:wgsymcut16}.


We observe five delay time peaks, which are Fano resonances resulting
from the closed cut billiard eigenstates 1a, 1b, 2a, 2b, and 3a shown
in Fig.~\ref{fig:bileigval}. Starting from the left, we label these
delay time peaks in the same manner. For cut size $\omega=20^\circ$,
we designate the delay time peaks, going from left to right, as 1b,
2a, 2b, 3a, and 3b. 1a is the resonance chain in the upper left
corner. The first two delay time peaks, 1a and 1b, are Fano resonances
resulting from the circle eigenstate with $(l=3,n=2)$, the following
two from $(l=6,n=1)$, and the last two, 3a and 3b, from
$(l=1,n=3)$. 

Again, we observe an avoided crossing between 2b and 3a, and the two
delay time peaks 2a and 2b actually cross. Like for the closed cavity,
the crossing is possible due to the parity symmetry of the system,
which has been preserved by adding the leads symmetrically with
respect to the cut. Since 2a and 2b have different parities, these
eigenvalue chains are allowed to cross.

The symmetric placement of leads has not broken any additional
symmetries. Only the energies and cut sizes at which the avoided
crossing and actual crossing take place are slightly shifted from the
case of the closed cut billiard.

\subsection{Asymmetric placement of leads and cut}
\label{cgasym}

In Fig.~\ref{fig:wgasymcut16} we show the same Fano resonances as in
Fig.~\ref{fig:wgsymcut16}, but for asymmetric placement of leads. Now
we place the leads at $\theta_1=0^\circ$ and $\theta_2=166^\circ$ and
the cut at $\zeta=225^\circ$.


The crossing of resonances 2a and 2b, which occurred for the symmetric
leads and for the closed cut-circle billiard, has now become an
avoided crossing. (In fact we now have a sequence of two closely
spaced avoided crossings, similar to the case studied in
Ref.~\CITE{todd}.) Here, the parity symmetry of the closed system is
destroyed by placing the leads asymmetrically. Therefore, in the
asymmetric open system, we no longer have two separate classes of
states (namely states of even and odd parity) and the crossing, which
was possible in the closed system and in the symmetric open system, is
avoided here.

\subsection{Cavity wavefunctions}
\label{cgstates}

We now look at the distribution of electron probability in the cavity
at the resonance energies. This is shown in Fig.~\ref{fig:resstates}
for the asymmetric system. A three-way avoided crossing has mixed the
resonance states, and has destroyed the identity they had for smaller
cut size. The upper row displays the four states before the avoided
crossing, which still reflect the quantum numbers of the associated
circle states very well. For the middle row, the character of states
2a, 2b, and 3a has been changed. Only the resonance state 1b, which
never comes really close to the crossing region, has preserved its
characteristic structure (two rings with six peaks each) quite
well. In the bottom row, after the second avoided crossing, 2a and 2b
have mixed even further.


%
%
\section{Conclusion}

\begin{figure}[h]
\begin{center}
\includegraphics[scale=1.0]{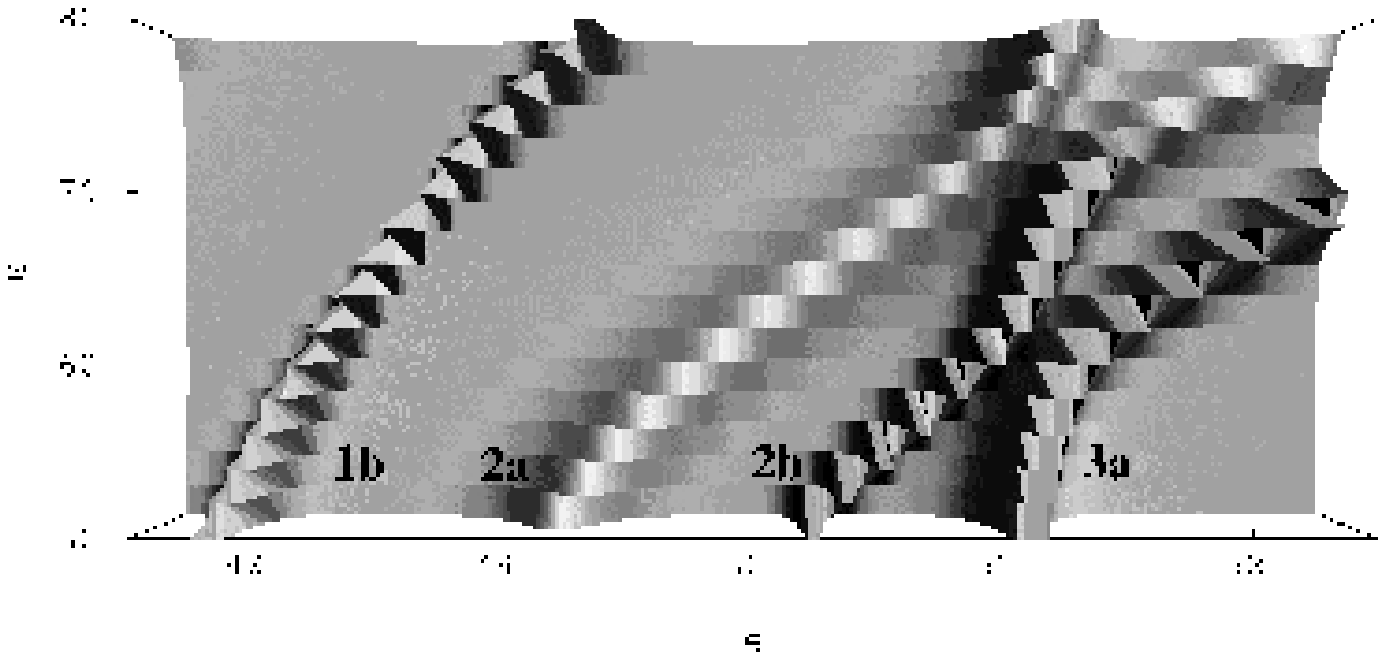}
\caption{Spectrum of the sum of partial delay times for the energy
interval $47.5{\leq}\epsilon{\leq}52.5$ and cut size interval
$52^\circ{\leq}\omega{\leq}80^\circ$. The lead openings are
$\Delta\theta_1=\Delta\theta_2=16^\circ$. Lead and cut positions are
$\theta_1=0^\circ$, $\theta_2=166^\circ$, and $\zeta=225^\circ$.
\label{fig:wgasymcut16}}
\end{center}
\end{figure}

\begin{figure}[h]
\begin{center}
\includegraphics[scale=1.0]{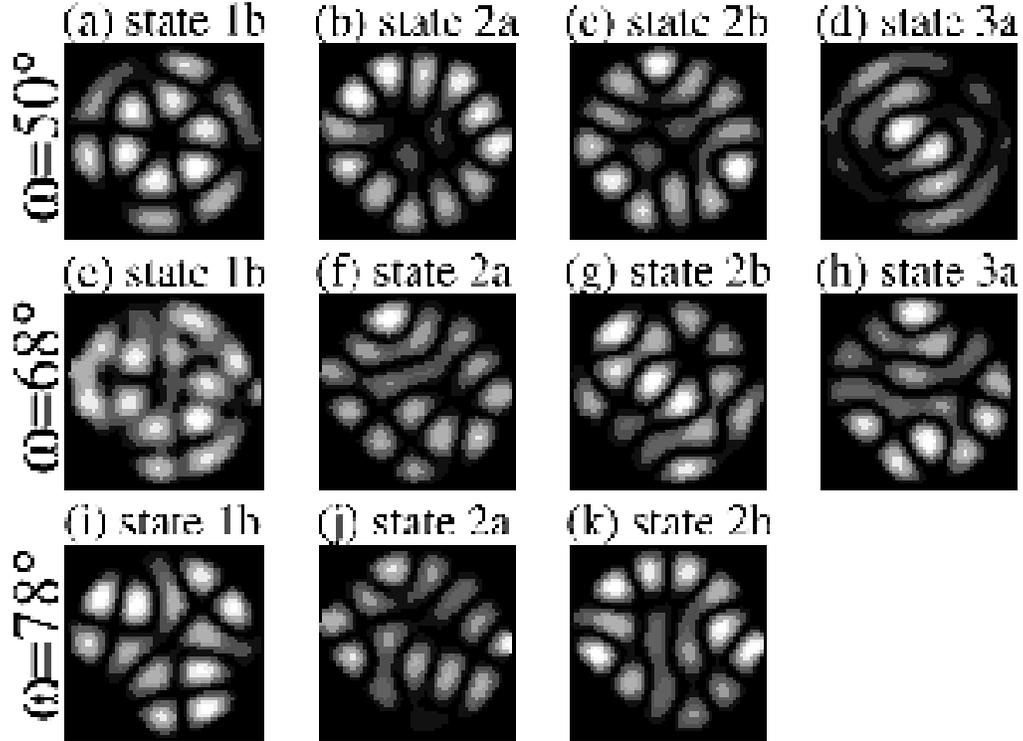}
\caption{Electron probability distribution in the waveguide cavity at
the resonance energy for Fano resonances shown in
Fig.~\ref{fig:wgasymcut16}. The lead openings are
$\Delta\theta_1=\Delta\theta_2=16^\circ$. Lead and cut positions are
$\theta_1=0^\circ$, $\theta_2=166^\circ$, and $\zeta=225^\circ$. The
top row has $\omega=50^\circ$ and consists of (a) State 1b for
$\epsilon=47.8$, (b) State 2a for $\epsilon=49.2$, (c) State 2b for
$\epsilon=50.2$, and (d) State 3a for $\epsilon=51.1$.  The middle row
has $\omega=68^\circ$ and consists of (e) State 1b for
$\epsilon=48.5$, (f) State 2a for $\epsilon=50.7$, (g) State 2b for
$\epsilon=51.3$, and (h) State 3a for $\epsilon=52.3$.  The bottom row
has $\omega=78^\circ$ and consists of (i) State 1b for
$\epsilon=49.2$, (j) State 2a for $\epsilon=50.7$, (k) State 2b for
$\epsilon=52.2$.
\label{fig:resstates}}
\end{center}
\end{figure}

In this paper, we investigated a waveguide with a cut-circle cavity by
calculating conductances and Wigner delay times.  We attached conic
leads to the cavity in two ways, symmetrically and asymmetrically.  We
observed two kinds of avoided crossings when the leads were placed
asymmetrically, breaking the parity symmetry of the closed cavity. One
avoided crossing was due to the chaos of the closed cavity, i.e., due
to level repulsions between states in the same symmetry class, which
are known to occur for fully chaotic systems. The other avoided
crossing was due to the breaking of the discrete symmetry (the parity
in our case) by placing leads asymmetrically (see
Fig.~\ref{fig:wgasymcut16}).

There is now great interest in exploring the effects of underlying
chaos on the dynamics of open quantum systems. However, open quantum
systems must make contact with the outside world. This contact itself
may induce some of the signatures of chaos. This fact has also been
studied extensively by Jung and Seligman \cite{jung} for classical
scattering systems. The avoided crossing between 2a and 2b, shown in
Fig.~\ref{fig:wgasymcut16}, provides another example of this effect.

\appendix

\section{Numerical method}
\label{sec:nummethod}
\subsection{Basic concept}
In order to obtain an explicit expression for the S-matrix, we need to
find the stationary solutions $\Psi$ of Eq.~(\ref{eq:schrod}). This
can only be done numerically for our system. For that purpose, we use
the boundary integral method described in Ref.~\CITE{frohne}. Below,
we will give an outline of this method adapted to our particular
problem.

The boundary integral method is based on the use of Green's Theorem,
which, for two functions $f$ and $g$, states that
\begin{equation}
\label{=GreenTh0}
\oint_\Gamma dl \left( f {\frac{\partial g}{\partial n} }
- {\frac{\partial f}{\partial n} }{g}\right)=
\int_A dA'({f}{\nabla}^2{g}-{g}{\nabla}^2{f})\:,
\end{equation}
if $\Gamma$ is a closed contour confining an area $A$, $l$ is the
longitudinal coordinate along $\Gamma$, and $n$ is the normal
coordinate.  For the boundary integral method, we choose for the
functions $f$ and $g$ the wavefunction $\Psi$ and a weight function
$\Phi$ that satisfies the Helmholtz equation
\begin{equation}
\label{=helmholtz_phi}
{\nabla}^2{\Phi}=-k^2\Phi\:.
\end{equation}
Then, since $\Psi$ is a solution of the Schr\"odinger equation
[Eq.~(\ref{eq:schrod})], i.e., $\Psi$ also satisfies
${\nabla}^2{\Psi}=-k^2\Psi$, the right hand side of
Eq.~(\ref{=GreenTh0}) is identically zero.  Therefore we get
\begin{equation}
\label{=GreenTh1}
\oint_\Gamma dl \left( \Psi {\frac{\partial \Phi}{\partial n}}
- {\frac{\partial \Psi}{\partial n}} {\Phi}
\right)=0\:
\end{equation}
as the basic equation of this numerical approach. (Note that the
difference between $\Psi$ and $\Phi$ is that there are no boundary
conditions for $\Phi$, whereas $\Psi$ has to satisfy the boundary
conditions imposed by the geometry of the waveguide.)

For our problem, we choose $\Gamma$ to follow the walls of the cavity.
For the places where the leads come in, $\Gamma$ is taken to be the
continuation of the circular wall segments in order to construct a
closed contour. Then, Eq.~(\ref{=GreenTh1}) allows us to obtain the
S-matrix and the probability distribution inside the cavity for any
incoming particle energy. This is done by inserting (approximate)
expansions of $\Psi$ with, say, $n_{\rm cutoff}$ yet unknown expansion
coefficients. As we need $n_{\rm cutoff}$ equations to solve for these
expansion coefficients, we apply Eq.~(\ref{=GreenTh1}) with
sufficiently many different weight functions $\Phi_n$
($n=1,\ldots,n_{\rm cutoff}$).

In principle, there are several choices possible for the $\Phi_n$'s,
as long as they satisfy Eq.~(\ref{=helmholtz_phi}). For numerical
convenience, we used the same functions as for the expansion of $\Psi$
inside the cavity, namely $\Phi_\gamma= f_\gamma$ with the
$f_\gamma$'s given by Eq.~(\ref{=basis-cavity}), where
$\gamma=-n_\gamma+1,\ldots,n_\gamma$ (i.e., $n_{\rm cutoff}= 2
n_\gamma$). For $n_\gamma$, we also used the same value as in
Sec.~\ref{cavstates}, $n_\gamma=90$, which allows us to determine
$n_{\rm cutoff}=180$ expansion coefficients.

\subsection{S-matrix}

\begin{figure}[h]
\begin{center}
\includegraphics[scale=1.0]{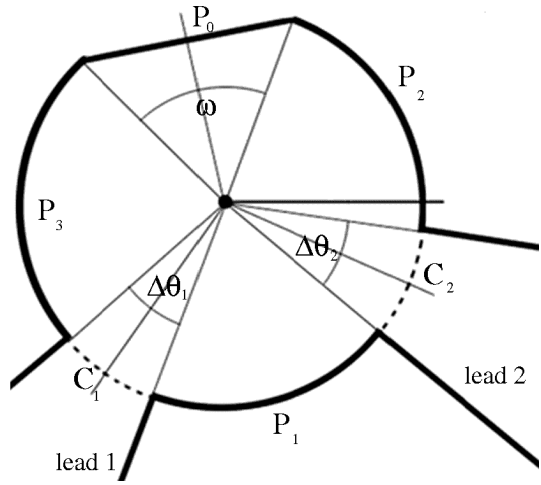}
\caption{Integration contour for the cut-circle waveguide, consisting of 
$C_1, C_2, P_0, P_1, P_2$, and $P_3$. \label{fig:geoapp}}
\end{center}
\end{figure}

In order to compute the S-matrix, we split up $\Gamma$ into parts
$C_1, C_2$ across the lead openings, $P_1, P_2, P_3$ along the
circular walls, and $P_0$ along the cut (see Fig.~\ref{fig:geoapp}).


We first regard a part $I_C$ of the integral in Eq.~(\ref{=GreenTh1}),
for which we only integrate over the lead openings $C_1$ and $C_2$
instead of $\Gamma$. Inserting for $\Phi$ the $f_\gamma$'s and for
$\Psi$ its expansion into lead channels from Eq.~(\ref{eq:leadexp1})
(for an electron incident in channel $\beta$ in lead~1), we obtain
\begin{eqnarray}
\label{=GreenTh2}
I_C &=& \int_{C_1} dl 
\left( \chi^{-}_{\beta} {\frac{\partial f_{\gamma}}{\partial n}}
- {\frac{\partial \chi^{-}_{\beta}}{\partial n}} f_{\gamma}\right)
+ \sum_\alpha r_{\alpha\beta} \int_{C_1} dl 
\left( \chi^{+}_{\alpha} {\frac{\partial f_{\gamma}}{\partial n}} 
- {\frac{\partial \chi^{+}_{\alpha}}{\partial n}} f_{\gamma}\right)
\nonumber \\
&&+ \sum_\alpha t_{\alpha\beta} \int_{C_2} dl 
\left( \chi^{+}_{\alpha} {\frac{\partial f_{\gamma}}{\partial n}} 
- {\frac{\partial \chi^{+}_{\alpha}}{\partial n}} f_{\gamma} \right) 
\nonumber \\ 
&=& (z_0^{1-})_{\gamma\beta}-(z_L^{1-})_{\gamma\beta} +
\sum_\alpha r_{\alpha\beta}
\left[ (z_0^{1+})_{\gamma\alpha}-(z_L^{1+})_{\gamma\alpha} \right] 
\nonumber \\
&&+ \sum_\alpha t_{\alpha\beta}
\left[ (z_0^{2+})_{\gamma\alpha}-(z_L^{2+})_{\gamma\alpha} \right] \:,
\end{eqnarray}
where we introduced the abbreviations
\begin{eqnarray}
\label{=lead-abbr}
(z^{i\pm}_L)_{\gamma\alpha} &\equiv& \int_{C_i} dl\:
\frac{\partial\chi^{i\pm}_\alpha}{\partial n} f_\gamma \:, \nonumber 
\\ (z^{i\pm}_0)_{\gamma\alpha} &\equiv& \int_{C_i} dl\:
\chi^{i\pm}_{\alpha} \frac{\partial f_\gamma} {\partial n} \:,
\end{eqnarray}
with $i=$1, 2 denoting lead 1 and 2. (For an electron incident in
lead~2, the expressions look similar.)

Along the walls of the cavity, the electron wavefunction is zero, but
it may have a finite slope perpendicular to the wall. We can expand
this slope in terms of a complete orthonormal basis along the
wall. Therefore, we can write
\begin{eqnarray}
\label{eq:wallexp}
\Psi_{\beta,P_i}&=&0 \:, \nonumber \\
\left( \frac{\partial \Psi_\beta}{\partial n} \right)_{P_i}&=&
\sum_\nu \:b^{(i)}_{\nu\beta} \;\xi^{(i)}_\nu (l),
\end{eqnarray}
with some set of basis functions $\xi^{(i)}_\nu$ in variable $l$. The
index $i=0$ is used to denote the cut, and $i=1,2,3$ denote the three
different circular wall segments.  We are free to choose any complete
basis, and for the circular wall segments we choose a Fourier basis,
\begin{equation}
\label{=basis-walls}
\xi^{(i)}_{\nu} (l)=
\sqrt{\frac{2}{\Delta_i}} \cos {\biggl[}\frac{\pi\nu \left
( l-{\delta}_i+\frac{\Delta_i}{2} \right)} {\Delta_i}{\biggr]} \:,
\end{equation}
where ${\delta}_i$ is the angular position of the center of wall
segment $i=1,2,3$, $\Delta_i$ is its opening angle, $\nu=0,1,\ldots$,
and ${\delta}_i-\frac{\Delta_i}{2}\leq l < {\delta}_i +
\frac{\Delta_i}{2}$. For the cut, we could use the Fourier basis,
too. However, as this results in very long computation times, a basis
of small triangles proved to be more efficient.  Thus, along the
straight wall we use
\begin{equation}
\label{=basis-cut-tri}
\xi^{(0)}_{\nu} (l)=\frac{2(n_{\nu_0}+1)}{l_0} \left\{
\begin{array} {lcr}
l-\frac{\nu}{n_{\nu_0}+1}\;l_0 & {\rm for}&
	\frac{\nu}{n_{\nu_0}+1}\; l_0 \leq l<
	\frac{\nu+1/2}{n_{\nu_0}+1}\; l_0\\
\frac{\nu+1}{n_{\nu_0}+1}\; l_0-l & {\rm for}&
	\frac{\nu+1/2}{n_{\nu_0}+1}\; l_0
	\leq l< \frac{\nu+1}{n_{\nu_0}+1}\; l_0\\
0 & {\rm else}&
\end{array} \right.\;,
\end{equation}
for $0 \leq l < l_0$, where $l_0=2R\sin (\omega/2)$ is the length of
the cut, $\omega$ is its opening angle, and $\nu=0,1, \ldots,
n_{\nu_0}$. Now we can write down the part $I_P$ of the integral in
Eq.~(\ref{=GreenTh1}), for which we only integrate over the wall
segments $P_0,\ldots,P_3$ instead of $\Gamma$. Inserting the
expansions from Eq.~(\ref{eq:wallexp}), we get
\begin{equation}
\label{=GreenTh3}
I_P= -\sum_{i=0}^3 \sum_{\nu} b^{(i)}_{\nu\beta} \int_{P_i} dl\:
\xi^{(i)}_{\nu} f_{\gamma} 
= -\sum_{i=0}^3 \sum_{\nu} b^{(i)}_{\nu\beta} ~ (x^i)_{\gamma\nu} \:,
\end{equation}
with the abbreviation
\begin{equation}
\label{=wall-abbr}
(x^i)_{\gamma\nu} \equiv \int_{P_i} dl\: \xi^{(i)}_\nu (l) f_\gamma \:.
\end{equation}

Now we can combine these results to obtain a matrix equation for the
unknown expansion coefficients, which can easily be solved
numerically. With the integrals $I_C$ and $I_P$,
Eq.~(\ref{=GreenTh1}) becomes $I_C+I_P=0$. Inserting
Eqs.~(\ref{=GreenTh2}) and~(\ref{=GreenTh3}) yields (after a slight
reordering of terms)
\begin{eqnarray}
\label{=GreenTh4}
\sum_\alpha 
\left[ (z_0^{1+})_{\gamma\alpha}-(z_L^{1+})_{\gamma\alpha} \right] 
r_{\alpha\beta} ~+~ \sum_\alpha 
\left[ (z_0^{2+})_{\gamma\alpha}-(z_L^{2+})_{\gamma\alpha} \right] 
t_{\alpha\beta} \nonumber \\ 
+\sum_{i=0}^3 \sum_{\nu} (-x^i)_{\gamma\nu} ~ b^{(i)}_{\nu\beta} 
~~=~~ - \left[ (z_0^{1-})_{\gamma\beta}-(z_L^{1-})_{\gamma\beta} 
\right] \:,
\end{eqnarray}
or, in compact matrix form:
\begin{equation}
\label{eq:frohne-matrix}
\left( \begin{array}{lr} {\bf Z}^+_0-{\bf Z}^+_L\;,&
	-{\bf X} \end{array}\right)
\left( \begin{array}{c} {\bf r}\\{\bf t}\\ {\bf b} 
        \end{array}\right)_\beta =
\left({\bf z}^{\;-}_L-{\bf z}^{\;-}_0\right)_\beta\:.
\end{equation}
The matrices ${\bf Z}^+_0$, ${\bf Z}^+_L$, and ${\bf X}$, as well as
the column matrix $\left({\bf z}^{-}_L-{\bf z} ^ {-}_0\right)_\beta$
can be computed numerically from Eqs.~(\ref{=lead-abbr})
and~(\ref{=wall-abbr}). Note that in order to include both leads
($i=1,2$) in ${\bf Z}^+_0$ and ${\bf Z}^+_L$, the first columns of
these matrices correspond to lead~1, and the columns for lead~2 are
simply appended. (Similarly, ${\bf X}$ is composed of the different
wall parts, $i=0,\ldots,3$, also columnwise.)

The quantity $\left(\begin{array}{c} {\bf r}\\ {\bf t}\\ {\bf b}
\end{array}\right) _\beta$ is a column matrix containing the unknown
transmission and reflection coefficients for a particle incident in
channel $\beta$ in lead 1, as well as the coefficients
$b^{(i)}_{\nu\beta}$, which describe the normal derivative of
$\Psi_\beta$ on the walls. As this is the quantity which we want to
solve for, we need to invert the matrix $\left( \begin{array}{lr} {\bf
Z}^+_0-{\bf Z}^+_L\;,& -{\bf X} \end{array}\right)$. Therefore, this
matrix has to be square. This can be achieved by using appropriate
cutoff numbers $n_{\alpha_i}$ ($i=1,2$) in the lead expansions
[Eq.~(\ref{eq:leadexp1})] and $n_{\nu_i}$ ($i=0,\ldots,3$) in the wall
expansions [Eq.~(\ref{eq:wallexp})]. Since the matrix consists of $2
n_\gamma$ rows, these numbers have to satisfy the relation
\begin{equation}
\label{=cutoffs}
\sum_{i=1}^2 n_{\alpha_i}+ \sum_{i=0}^3 (n_{\nu_i}+1) = 2 n_\gamma \:.
\end{equation}
(Note that $\nu=0,\ldots,n_{\nu_i}$, therefore we actually sum over
$n_{\nu_i}+1$ terms for the wall parts.) For best numerical accuracy,
the relations of the cutoff numbers are chosen approximately equal to
the length relations of the corresponding integration paths.

\subsection{Cavity probability amplitude}

Once the elements of the S-matrix have been obtained by the method
described above, it is an easy task to calculate the probability
amplitude for the electron state inside the cavity. We know that the
wavefunction, which, inside the cavity, is described by
Eq.~(\ref{eq:cavstate}), is zero along all the walls and that is
continuous at the interface between the leads and the cavity. Thus we
require that along the walls,
\begin{equation}
\label{eq:cavity1}
{\sum_\gamma}c_{\gamma\beta} f_{\gamma}(\rho,\varphi)=0\:,
\end{equation}
and along the interface $C_1$ and $C_2$ between the leads and cavity
\begin{equation}
\label{eq:cavity2}
{\sum_\gamma}c_{\gamma\beta}
f_{\gamma}(\rho,\varphi) = \Psi_{\beta}(\rho,\varphi)\:,
\end{equation}
where $\Psi_\beta$ (in the leads, for incoming mode $\beta$) is known
from the previous section.  It is straightforward to solve for the
$c_{\gamma\beta}$'s, e.g., by evaluating the above expressions at $2
n_\gamma$ different points on $\Gamma$, and inverting the resulting
system of linear equations.

\end{document}